\documentclass{PoS}
\usepackage{wrapfig}
\newcommand{\pT}{$p_{T}$}

\newcommand{\sNN}{$\sqrt{s_{_{NN}}}$}

\newcommand{\DirPho}{$\gamma_{\rm dir}$}
\newcommand{\piZro}{$\pi^{0}$}

\newcommand{\ptAssoc}{$p_{T}^{\mathrm{assoc}}$}
\newcommand{\ptTrig}{$p_{T}^{\mathrm{trig}}$}
\newcommand{\zT}{$z_{T}$}

\newcommand{\IAApiZro}{$I_{AA}^{\pi^{0}}$}
\newcommand{\IAAg}{$I_{AA}^{\gamma_{dir}}$}

\title{Jets and high-$p_{T}$ probes measured in the STAR experiment}

\ShortTitle{Jets and high-$p_{T}$ probes measured in the STAR experiment}

\author{{Nihar Ranjan Sahoo (for the STAR collaboration)}%\thanks{A footnote may follow.}\\
\\
       Texas A$\&$M University, Texas, USA\\\
        E-mail: \email{nihar@rcf.rhic.bnl.gov}}

\abstract{Hard probes created through large momentum transfers are used to study
  the properties of QCD matter created in heavy-ion  collisions, by
  comparing the measurements to those in p+p collisions.  Jets, and the
  "quenching" or suppression of jets in the medium created in heavy-ion
  collisions, are studied through various different observables.  We
  present the most recent measurements from $\sqrt{s_{NN}}$ = 200 GeV
  Au+Au collisions, with p+p collisions as the reference, by the STAR
  Collaboration.  The observables are semi-inclusive charged
	jets and  di-jet transverse momentum imbalance.  Additionally, correlation
  measurements of direct photon-hadron and neutral pion-hadron are
  presented and discussed.}

\FullConference{38th International Conference on High Energy Physics\\
		3-10 August 2016\\
		Chicago, USA}

\begin{document}

\section{Introduction}
Jets and high-\pT~ particles are produced on very short time scales ($\sim 0.1 \rm fm/c$) in collisions with large momentum transfer (\pT~$>  \rm Q_{0} \gg \Lambda_{QCD}$).
Hence they are considered good tomographic probes of the hot and dense QCD medium created in heavy-ion collisions. Over the last decade or so, many compelling measurements, such as the disappearance of away-side jets and high-\pT~suppression~\cite{STAR_jet_highpT}, di-jet suppression~\cite{STAR_Dijet} and high-\pT~suppression balanced by low \pT~enhancement in jet-hadron correlation~\cite{STAR_jetH} etc., contributed to our understanding of jet quenching in the medium created at RHIC. In these proceedings, I discuss three recent measurements in the STAR experiment: (i) Jet-like direct photon-hadron and $\pi^{0}$-hadron correlations, (ii) di-jet transverse momentum imbalance, and (iii) semi-inclusive recoil charged jets.

%%%%%%%%%%%%%%%%%%%%%% Figure -5 %%%%%%%%%%%%%%%%% 
%\begin{figure}[htbp]
%\begin{center}
\begin{wrapfigure}{r}{0.5\textwidth}
\vspace{-10pt}
\includegraphics[width=0.5\textwidth]{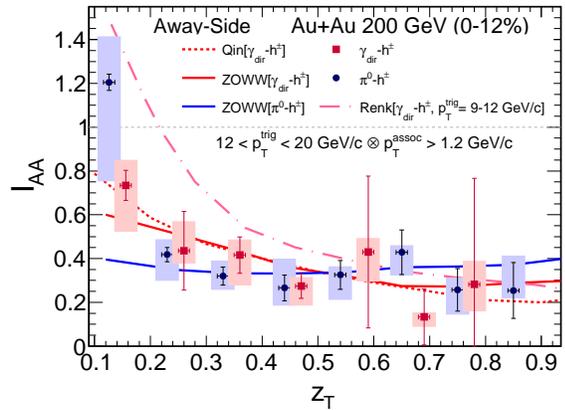}
\vspace{-6pt}
\caption{(Color online.) The  $I_{AA}$~for \DirPho~(red  squares) and \piZro~(blue circles) triggers are plotted as a function of \zT. The points for $I_{AA}$~for \DirPho~ are shifted by $+0.03$ in \zT~for visibility. The vertical line and shaded boxes represent statistical and systematic errors, respectively~\cite{STAR_GammaHadron_PLB}. The curves  represent theoretical model predictions~\cite{ZOWW,Wang,Qin,YAJEM}. }
\label{Fig1}
%\vspace{-5pt}
%\end{center}
%\end{figure}
\end{wrapfigure}

\section{Jet-like direct photon-hadron and $\pi^{0}$-hadron correlation}
\begin{figure}[htbp]
\begin{center}
\includegraphics[width=0.5\textwidth]{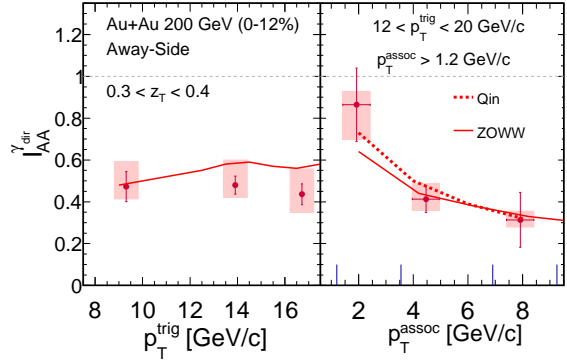}
\caption{(Color online.) \IAAg~are plotted as a function of \ptTrig~(left panel) and \ptAssoc~(right panel). The vertical line and shaded boxes represent statistical and systematic errors, respectively. The curves  represent theoretical model predictions~\cite{ZOWW,Wang,Qin}. }
\label{Fig2}
\end{center}
\end{figure}

The motivation for jet-like direct photon-hadron and $\pi^{0}$-hadron correlation studies is to understand the flavor and path length dependence of parton energy loss in the hot and dense medium~\cite{STAR_GammaHadron_PLB}. In this analysis, the  triggered  \DirPho~and \piZro~are selected with 12$<$ \ptTrig $<$ 20 GeV/c and charged tracks with 1.2 GeV/c$<$ \ptAssoc~in order to attain low \zT (=$p_{T}^{\rm assoc}/p_{T}^{\rm trig}$) values down to 0.1. A detailed discussion and analysis techniques can be found in the Ref.~\cite{STAR_GammaHadron_PLB}. The suppression of these jet-like yields in central Au+Au collisions is then quantified by comparing to the per-trigger yields measured in p+p collisions, denoting the ratio of integrated yields $I_{AA}$. The away-side medium modification for \DirPho~(\IAAg) and \piZro~ (\IAApiZro) triggers are shown as a function of \zT~in Fig.~\ref{Fig1}.
The away side $I_{AA}$  for both triggers has a systematic trend to lower values with increasing \zT~though not significant within uncertainties. This observation is somewhat more significant when $I_{AA}$ is plotted as a function of \ptAssoc~in Fig.~\ref{Fig2} (right panel). The expected difference between \IAAg~and \IAApiZro~triggers as in models~\cite{ZOWW,Wang} at low \zT~is difficult to observe because of large uncertainties in the data. \IAAg ~is plotted for three \ptTrig~bins ranging from 8 to 20 GeV/c for $0.3 <$ \zT~$< 0.4$ in Fig~\ref{Fig2} (left panel). It is found that \IAAg~is insensitive to the \DirPho-trigger energy in this range at RHIC energy. %An enhancement of per \DirPho-trigger away-side associated yield within the full integration window of $|\Delta\phi - \pi| < 1.4$ over a smaller window of $|\Delta\phi - \pi| < 0.6$, is seen at low \zT. A detailed discussion and corresponding figure can be found in Ref~\cite{STAR_GammaHadron_PLB}.  %This observation concludes that the lost energy reappears predominantly at low \pT ~( \pT $< $2 GeV/c), regardless of the trigger \pT.
Further understanding on the redistribution of lost energy in heavy-ion collisions can be explored by measuring the distribution of fully reconstructed recoil jets with respect to a \DirPho-trigger. Such a measurement of charged and full jets is underway in the STAR experiment.

%%%%%%%%%%%%%%%%%%%%%%%%%%%%%%%%
%\end{wrapfigure}
%%%%%%%%%%%%%%%%%%%%%%%%%%%%%%%%

%%%%%%%%%%%%%%%%%%%%%%%%%%%%%%%%%%%%
\section{Semi-inclusive recoil charged jets}
%%%%%%%%%%%%%%%%%%%%%%%%%%%%%%%%
\begin{figure}[htbp]
\begin{center}
\includegraphics[width=0.45\textwidth]{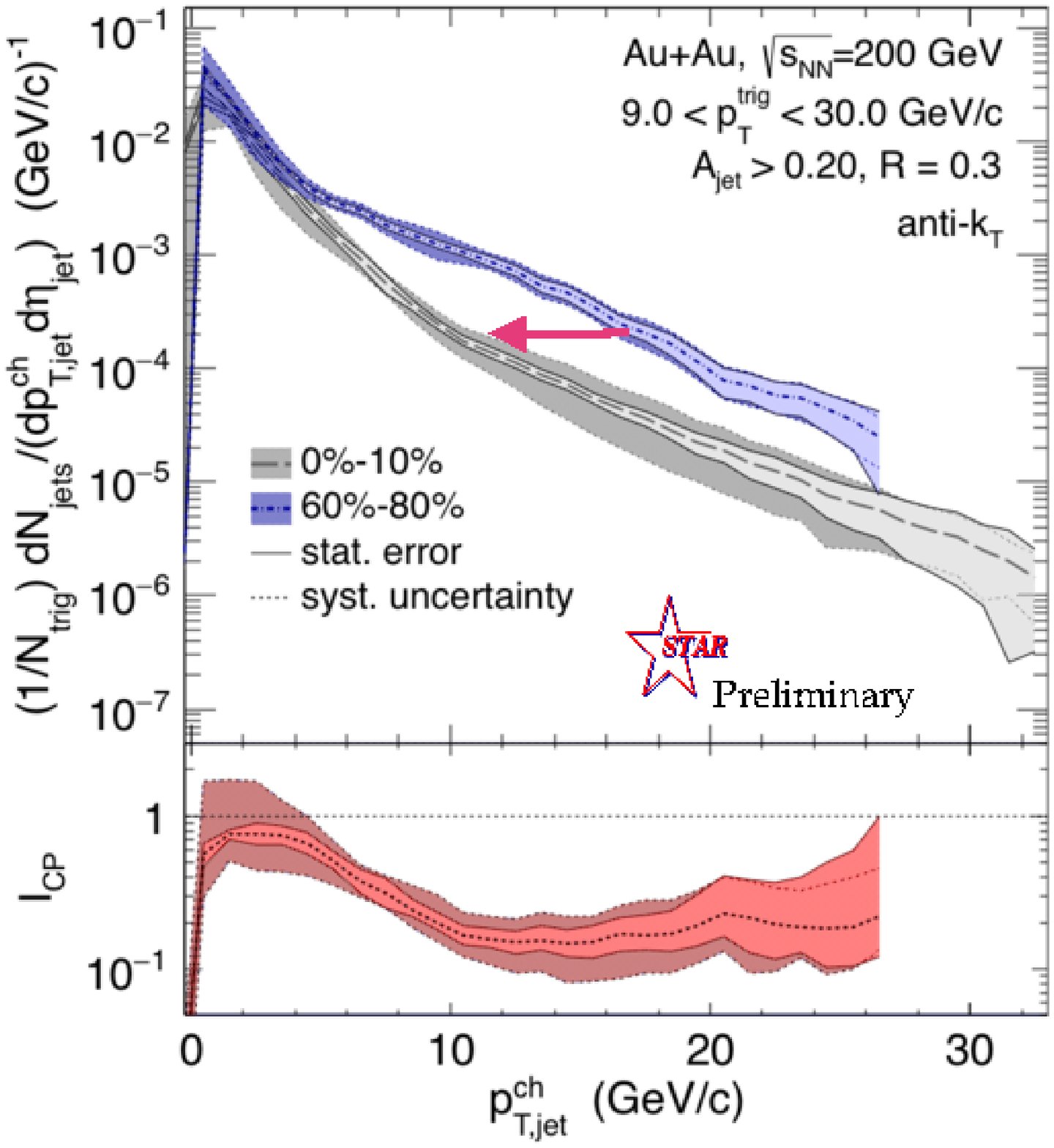}
\includegraphics[width=0.45\textwidth]{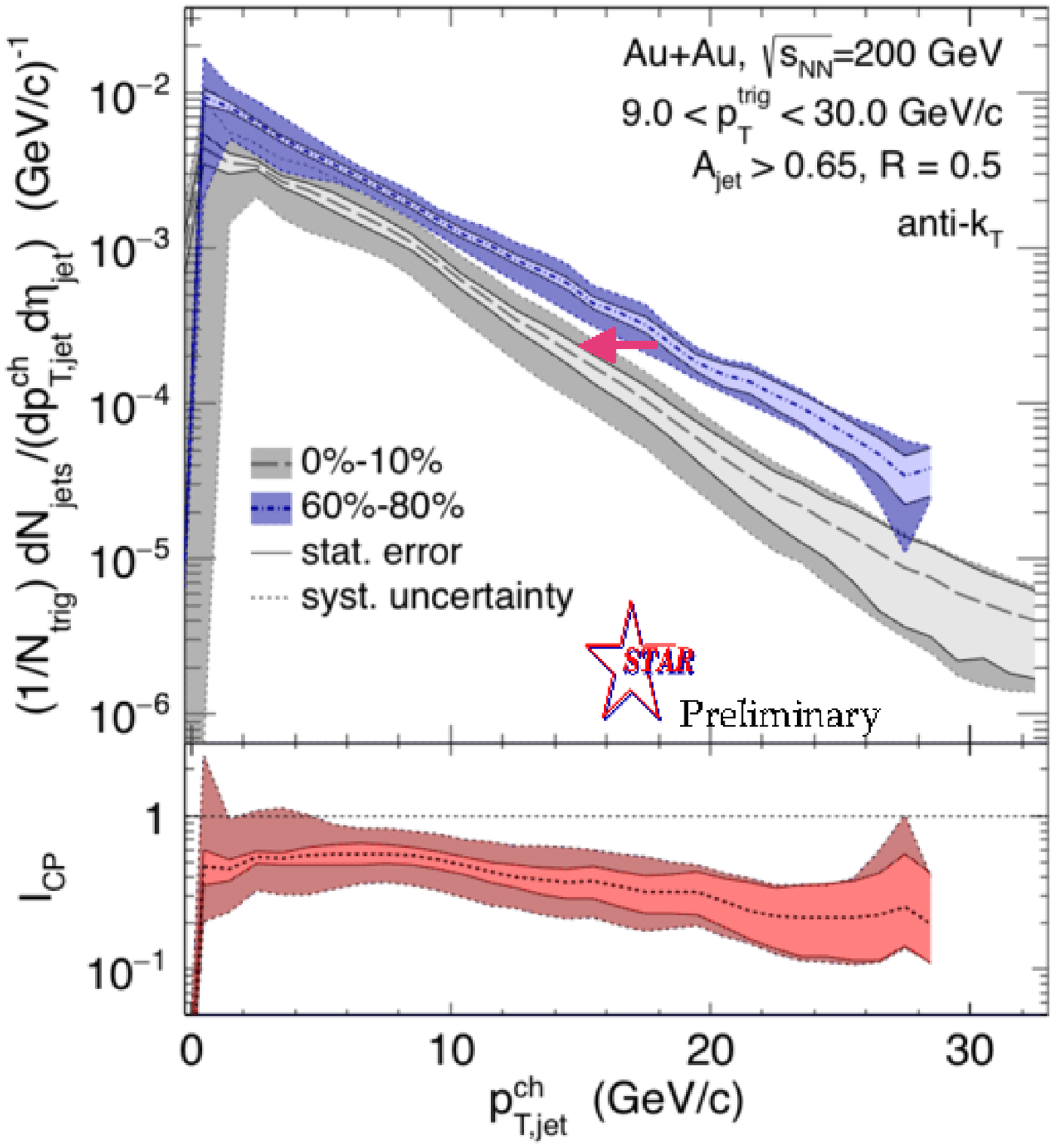}
\caption{(Color online.) Upper panels: Corrected charged recoil jet $p_{T,jet}^{ch} $ distributions for peripheral and central Au+Au collisions for R=0.3 (left) and R=0.5 (right). Arrow represents level of horizontal shift in $p_{T,jet}^{ch} $ spectra (guide to eyes)~\cite{STAR_hjet}. Lower panels: $I_{CP}$ for R=0.3 (left) and R=0.5 (right). }
\label{Fig31}
\end{center}
\end{figure}
%\end{wrapfigure}
%%%%%%%%%%%%%%%%%%%%%%%%%%%%%%%%

A new jet measurement performed in  the STAR experiment is the semi-inclusive charged jet spectrum on the recoil side of a high-\pT~charged-hadron trigger.
The reconstructed charged recoil jets are termed as semi-inclusive, since the triggered hadron \pT~is not inclusive (within 9 $<$ \ptTrig$<$ 30 GeV/c). This type of measurement is very challenging owing to the high-multiplicity environment and underlying background fluctuations in heavy-ion collisions. A novel mixed-event technique was used for correcting uncorrelated jet background from the reconstructed jets by a statistical subtraction method~\cite{ALICE,STAR_hjet}.  One trigger hadron is selected randomly in the above \pT~range and charged jets (consisting of charged tracks with \pT~> 0.2 GeV/c) are reconstructed using the anti-$k_{T}$ algorithm for a given resolution parameter (R = 0.3 and 0.5 for these results). The recoil jet acceptance is  in $|\pi - \Delta \phi| < \pi/4$. The estimated background energy density ($\rho$) scaled by jet area ($A$) is subtracted from each reconstructed jet raw transverse momentum ($p_{T,jet}^{raw}$),  $p_{T,jet}^{reco} =  p_{T,jet}^{raw} - \rho A$. This reconstructed jet $p_{T,jet}^{reco}$ spectrum is then corrected by subtracting that of mixed-events.  This raw correlated distribution is finally corrected by an unfolding procedure for instrumental effects and \pT-smearing due to the background. The upper panels of Fig.~\ref{Fig31} show the semi-inclusive corrected and recoil charged jet transverse momentum ($p_{T,jet}^{ch}$) spectra for peripheral and central Au+Au collisions for R=0.3 and 0.5. Significant suppression in central vs. peripheral, via the medium modification, $I_{CP}$, is observed for $p_{T,jet}^{ch} > 10$ GeV/c in case of R=0.3 and R=0.5. The horizontal shift in $p_{T,jet}^{ch}$ spectra in central compared with peripheral for R=0.3 indicates that the jet energy is transported out of the cone due to the {\it jet-quenching effect}. This horizontal shift is $-2.3 \pm 0.2$ GeV/c for R=0.5 and $-5.0\pm 0.5$ GeV/c for R=0.3 with $p_{T,jet}^{ch} > 10$.

\section{Di-jet transverse momentum imbalance}
%Di-jet measurement has been performed in the STAR experiment to understand the contribution of soft particles is the di-jet transverse momentum (\pT) imbalance. 

Di-jet measurement has been performed in the STAR experiment to understand the emission of soft particles with respect to the di-jet axis by measuring the di-jet transverse momentum (\pT) imbalance. The di-jet \pT~imbalance observable is defined as $A_{J} = (p_{T,lead} - p_{T,sublead} )/(p_{T,lead} + p_{T,sublead} )$. Where $p_{T,lead}$ and $p_{T,sublead}$ are the \pT~of the leading and sub-leading jets, respectively. Events were required to have a high tower trigger (HT) with an uncorrected transverse energy of $E_{T} > $ 5.4 GeV in the barrel electromagnetic calorimeter (BEMC) towers. In these HT events, $A_{J}$ is calculated using $p_{T,lead} > 20 $ GeV/c and $p_{T,sublead} > 10 $ GeV/c with $|\phi_{lead} - \phi_{sublead} - \pi | <$ 0.4. Full jets are reconstructed using charged tracks measured in the TPC and neutral tracks information recored in the BEMC using the anti-$k_{T}$ algorithm~\cite{antikt,FASTJET}. Details of the technique used in this analysis can be found in Ref.~\cite{STAR_dijetImb, STAR_kk}.

The upper panel of Fig.~\ref{Fig4} shows the normalized  distributions of $A_{J}$ for R=0.4 in Au+Au HT events compared with p+p HT $\oplus$ Au+Au MB events (events of p+p HT embedded into Au+Au 0-20$\%$ central events of minimum bias data sample) for constituents \pT ~$> 2 $ GeV/c.  It is observed that di-jets in Au+Au HT are significantly imbalanced compared with p+p HT $\oplus$ Au+Au MB events. This behavior is further studied by including soft particles \pT ~$> $0.2 GeV/c in jet reconstruction and then performing a geometrical matching ($\Delta R = \sqrt{\Delta \phi^{2} + \Delta \eta^{2} } < R$) with the initial hardcore di-jets. The di-jet imbalance is restored by including soft particles for jet cone parameter R=0.4. A similar study is also performed using R=0.2, and the $A_{J}$ distributions are shown in the lower panel of Fig.~\ref{Fig4}. It shows that the di-jet \pT~imbalance can not be restored including soft particles for R=0.2. The above observations indicate that the studied selection of "hard core" di-jets clearly experiences medium modification,
but in contrast to corresponding LHC measurements, the redistributed energy is still contained within the original $R = 0.4$ cone.
With a smaller cone size, balance cannot be recovered, suggestive of broadening of the jet structure compared with p+p collisions.

%This indicates that, at RHIC, the energy lost by di-jets ($p_{T,lead} > 20 $ GeV/c and $p_{T,sublead} > 10 $ ) with  interacting in the medium reappears as soft particles by broadening the jet structure compared with p+p collisions.

 %%%%%%%%%%%%%%%%%%%%%%%%%%%%%%%%

\begin{figure}[htbp]
\vspace{20pt}
\begin{center}
%\vspace{-40pt}
\includegraphics[width=0.7\textwidth]{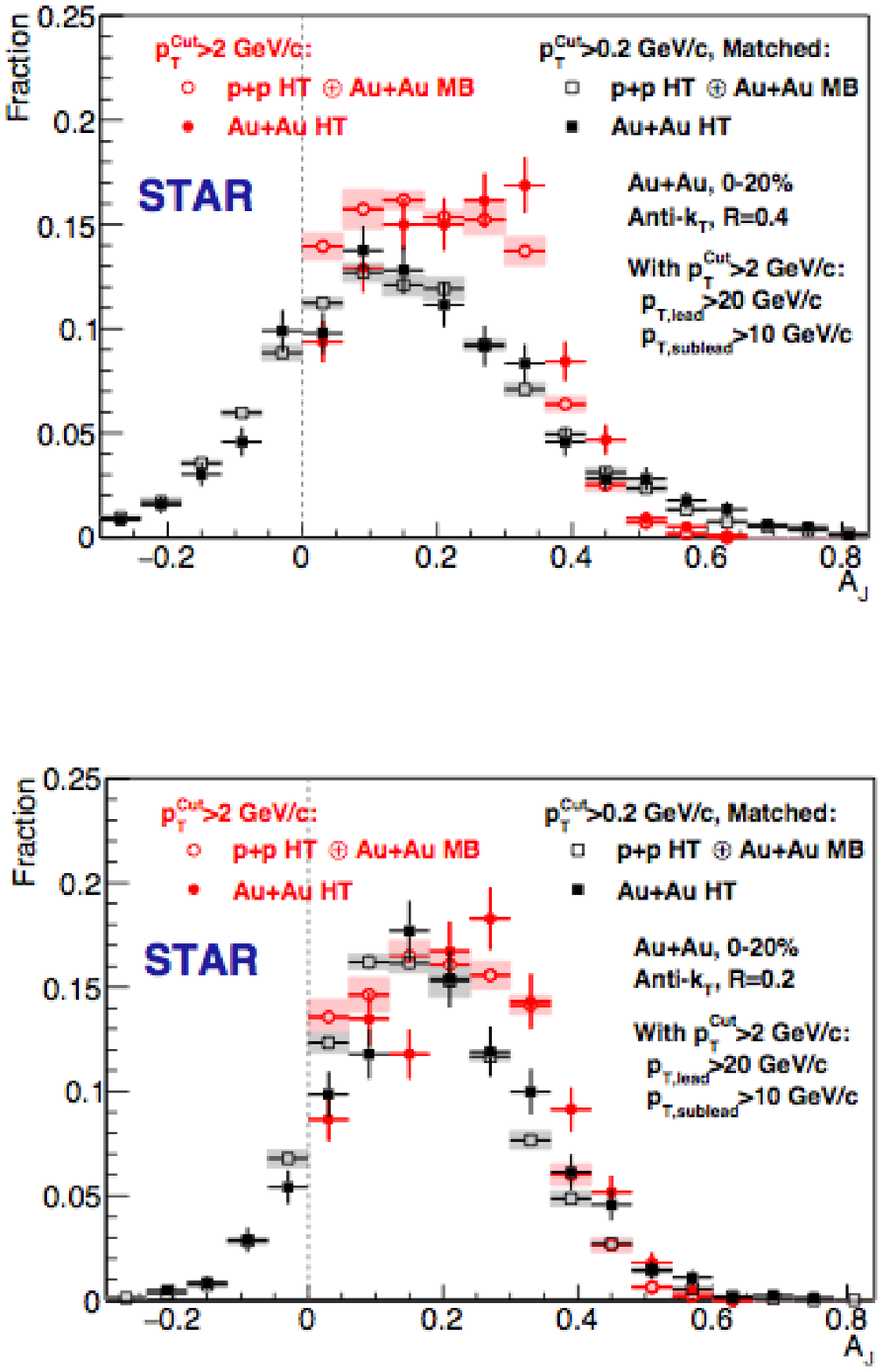}
%\vspace{-70pt}
%\includegraphics[width=0.5\textwidth]{R02}
 \label{Fig4}
\end{center}
\end{figure}
 \begin{figure}[htbp]
\vspace{-10pt}
\begin{center}
%%%%%%%%
%\vspace{-50pt}

\caption{(Color online.) Normalized $A_{J}$ distributions for
Au+Au HT data (filled symbols) and p + p HT $\oplus$ Au+Au
MB (open symbols). The red circles points are for jets
found using only constituents with $p_{T}^{\rm{Cut}} >$ 2 GeV/c and the
black squares for matched jets found using constituents  with $p_{T}^{\rm Cut} >$ 0.2 GeV/c~\cite{STAR_dijetImb, STAR_kk}. Upper panel: for R = 0.4. Lower panel: for R = 0.2.}
\label{Fig4}
\end{center}
\end{figure}

%%%%%%%%%%%%%%%%%%%%%%%%%%%%%%%%
 
\section{Summary}
The STAR experiment recently measured the following three jet observable to study the hot and dense matter created at RHIC.  
	\begin{itemize}
\item Jet-like direct photon-hadron and $\pi^{0}$-hadron correlations: Both \IAAg~and \IAApiZro ~show similar levels of suppression. The expected differences due to the color factor and path length dependence are not observed within current experimental
uncertainties. At top RHIC energy, no \DirPho-trigger energy dependence is observed on the suppression of away-side yields in the range of 8$< $ \ptTrig $<$ 20 GeV/c. The lost energy reappears predominantly at low \pT ~(\pT $< $2 GeV/c), regardless of the trigger \pT~of \DirPho.

\item Semi-inclusive recoil charged jets: A  novel mixed-event method was developed to correct the uncorrelated fake jets contribution in heavy-ion collisions in the STAR experiment. After this correction, the semi-inclusive recoil charged-jets spectra of a high-\pT~hadron trigger show $\sim$80\% suppression in recoil jet \pT~ in central collisions with respect to peripheral collisions with R=0.3. A significant horizontal shift in the recoil jet \pT~spectra in central collisions with respect to peripheral collisions at R=0.3 compared with that at R=0.5 indicates that a comparatively wider jet cone is the consequence of jet-quenching in heavy-ion collisions.

\item Di-jet transverse momentum imbalance: A significant di-jet imbalance is observed in Au+Au collisions in comparison with the p+p reference for the jet resolution parameter R=0.4 including constituent particles with \pT $>$ 2 GeV/c. When including softer particles (with \pT~> 0.2 GeV/c), the balance is
restored to the level of the embedded p+p reference, indicating that redistributed energy is still contained within the original R = 0.4 cone, though not within a smaller jet resolution parameter of R=0.2. It indicates that the energy loss in di-jet events can not be  recovered within a narrow jet cone in heavy-ion collisions at \sNN ~= 200 GeV for this particular selection of di-jets.
\end{itemize}

 Beside these measurements, new jet measurements  like neutral triggered jets, soft drop grooming in jet etc., are ongoing in the STAR experiments to study the QCD medium.

\end{document}